\documentstyle[12pt]{article}
\begin{document}
\newcommand{\eqn}[1]{(\ref{#1})}
\renewcommand{\a}{\alpha}
\renewcommand{\b}{\beta}
\renewcommand{\c}{\gamma}
\renewcommand{\d}{\delta}
\newcommand{\th}{\theta}
\newcommand{\TH}{\Theta}
\newcommand{\pa}{\partial}
\newcommand{\g}{\gamma}
\newcommand{\G}{\Gamma}
\newcommand{\A}{\Alpha}
\newcommand{\B}{\Beta}
\newcommand{\D}{\Delta}
\newcommand{\e}{\epsilon}
\newcommand{\E}{\Epsilon}
\newcommand{\z}{\zeta}
\newcommand{\Z}{\Zeta}
\newcommand{\k}{\kappa}
\newcommand{\K}{\Kappa}
\renewcommand{\l}{\lambda}
\renewcommand{\L}{\Lambda}
\newcommand{\m}{\mu}
\newcommand{\M}{\Mu}
\newcommand{\n}{\nu}
\newcommand{\N}{\Nu}
\newcommand{\x}{\chi}
\newcommand{\X}{\Chi}
\newcommand{\p}{\pi}
\newcommand{\r}{\rho}
\newcommand{\R}{\Rho}
\newcommand{\s}{\sigma}
\renewcommand{\S}{\Sigma}
\renewcommand{\t}{\tau}
\newcommand{\T}{\Tau}
\newcommand{\y}{\upsilon}
\newcommand{\Y}{\upsilon}
\renewcommand{\o}{\omega}
\newcommand{\q}{\theta}
\newcommand{\h}{\eta}
\def\ft#1#2{{\textstyle{{#1}\over{#2}}}}
\def\del{\partial}
\def\modb{\bar \Phi}
\def\mat{C}
\def\matb{\bar C}
\def\mpl{M_{\rm Pl}}

\newcommand{\st}{\scriptstyle}
\newcommand{\sst}{\scriptscriptstyle}
\newcommand{\mco}{\multicolumn}
\newcommand{\epp}{\epsilon^{\prime}}
\newcommand{\vep}{\varepsilon}
\newcommand{\ra}{\rightarrow}
\newcommand{\ab}{\bar{\alpha}}
\def\be{\begin{equation}}
\def\ee{\end{equation}}
\def\bea{\begin{eqnarray}}
\def\eea{\end{eqnarray}}

\renewcommand{\section}[1]{\addtocounter{section}{1}
\vspace{5mm} \par \noindent
  {\bf \thesection . #1}\setcounter{subsection}{0}
  \par
   \vspace{2mm} } 
\newcommand{\sectionsub}[1]{\addtocounter{section}{1}
\vspace{5mm} \par \noindent
  {\bf \thesection . #1}\setcounter{subsection}{0}\par}
\renewcommand{\subsection}[1]{\addtocounter{subsection}{1}
\vspace{2.5mm}\par\noindent {\em \thesubsection . #1}\par
 \vspace{0.5mm} }
\renewcommand{\thebibliography}[1]{ {\vspace{5mm}\par \noindent{\bf 
References}\par \vspace{2mm}} 
\list
 {\arabic{enumi}.}{\settowidth\labelwidth{[#1]}\leftmargin\labelwidth
 \advance\leftmargin\labelsep\addtolength{\topsep}{-4em}
 \usecounter{enumi}}
 \def\newblock{\hskip .11em plus .33em minus .07em}
 \sloppy\clubpenalty4000\widowpenalty4000
 \sfcode`\.=1000\relax \setlength{\itemsep}{-0.4em} }
 
\begin{titlepage}
\begin{flushright} THU-96/14\\
hep-th/9603191
\end{flushright}
\vfill
\begin{center}
{\large\bf N=2 SYMPLECTIC REPARAMETRIZATIONS IN A CHIRAL 
BACKGROUND${}^\dagger$ }   \\ 
\vskip 7.mm
{B. de Wit }\\
\vskip 0.1cm
{\em Institute for Theoretical Physics, Utrecht University}\\
{\em Princetonplein 5, 3508 TA Utrecht, The Netherlands} \\[5mm]
\end{center}
\vfill
 
\begin{center}
{\bf ABSTRACT}
\end{center}
\begin{quote}
We study the symplectic reparametrizations that are possible 
for theories of $N=2$ supersymmetric vector multiplets in the 
presence of a chiral background and discuss some of their 
consequences. One of them concerns an anomaly 
arising from a conflict between symplectic covariance and 
holomorphy.   %
\vfill      \hrule width 5.cm
\vskip 2.mm
{\small\small
\noindent $^\dagger$ Based on a talk given at the Workshop on Recent 
Developments in Theoretical Physics, ``STU-Dualities and 
Nonperturbative Phenomena in Superstrings and Supergravity'', 
Cern Theory Division, 29 November - 1 December, 1995; to appear 
in Fortschritte der Physik.}  
\end{quote}
\begin{flushleft}  
March 1996
\end{flushleft}
\end{titlepage}

\vspace{4mm} 
\begin{center}
{\bf N=2 SYMPLECTIC REPARAMETRIZATIONS IN A CHIRAL BACKGROUND }
\vspace{1.4cm}
 
B.~DE~WIT \\
{\em Institute for Theoretical Physics, Utrecht University} \\
{\em Princetonplein 5, 3508 TA Utrecht, The Netherlands} \\
\end{center}             
\centerline{\small ABSTRACT}
\vspace{- 2 mm}  
\begin{quote}\small
We study the symplectic reparametrizations that are possible 
for theories of $N=2$ supersymmetric vector multiplets in the 
presence of a chiral background and discuss some of their 
consequences. One of them concerns an anomaly 
arising from a conflict between symplectic covariance and 
holomorphy.   %
\end{quote}
\addtocounter{section}{1}
\par \noindent
  {\bf \thesection . Introduction}
  \par
   \vspace{2mm} 

\noindent
Theories of abelian $N=2$ vector multiplets 
transform systematically under duality transformations: 
transformations acting on the (abelian) field strengths which  
rotate the combined field equations  
and Bianchi identities by means of a real symplectic matrix. This was 
first exploited for pure $N=2$  
supergravity \cite{SG}. For generic $N=2$ vector supermultiplets it 
was discovered \cite{DWVP} that these 
transformations rotate the scalar fields $X^I$ and the derivatives 
$F_I$ of the holomorphic function $F(X)$ that encodes the 
Lagrangian, by means of the same $Sp(2n+2;{\bf R})$ transformation, 
where $n$ denotes the number of vector multiplets\footnote{%
   Not counting the graviphoton of $N=2$ supergravity. In the 
   rigid case the symplectic matrix is only $2n$-dimensional.}. %
Initially the emphasis was on invariances of the equations of 
motion. The fact that the scalars in supergravity often 
parametrize an homogeneous  
space whose transitive isometries are realized through duality 
transformations, enables one to conveniently controll the 
nonpolynomial dependence on the scalar fields. Later it was 
realized that these transformations can also be used to 
reparametrize the theory in terms of a different function $\tilde 
F(\tilde X)$ \cite{CecFerGir}. For the subgroup of the 
symplectic group corresponding to an invariance of the equations 
of motion, the function $F$ will remain the same. 

The same symplectic reparametrizations emerged in the context  of 
type-II string compactifications on Calabi-Yau manifolds, where 
$(X^I,F_J)$ can be associated with the periods of the $(3,0)$ form of the 
Calabi-Yau three-fold. These periods transform under symplectic 
rotations induced by changes in the 
corresponding homology basis \cite{special,Cand}. The scalar 
sector of the vector multiplets, which in this application 
corresponds to (part of) the moduli space of the Calabi-Yau 
manifolds, are therefore subject to the same symplectic 
transformations.  

More recently symplectic reparametrizations were exploited by 
Seiberg and Witten  
\cite{SW}, and later by others \cite{Lerche}, in 
obtaining exact solutions of low-energy effective actions for 
$N=2$ supersymmetric Yang-Mills theory. Singularities in these 
effective actions signal their breakdown  
due to the emergence of massless states corresponding to 
monopoles and dyons.  Although these states are the result of 
nonperturbative dynamics, they are nevertheless accessible 
because at these points one conveniently converts to an 
alternative dual formulation, in which local field theory is 
again applicable. The same ideas have been used in the description 
of the effective action of heterotic $N=2$ compactifications 
[8-10].

Both in the context of Calabi-Yau manifolds and in the 
nonperturbative sector of supersymmetric Yang-Mills theories, the 
symplectic transformations are usually restricted to a discrete 
subgroup. This can be understood from the fact that they are 
related to changes of the homology basis, the periodicity of the 
generalized $\theta$-angles and/or the invariant rotations of the 
lattice of electric-magnetic charges. 

In this talk we discuss a number of features related to the 
symplectic reparametrizations in the presence of a chiral 
background. As some of this material has 
already been covered elsewhere \cite{buckow}, we will mainly 
summarize some of the results and clarify specific points.  

\section{Symplectic reparametrizations}
\noindent{}The actions we use are based on $N=2$ chiral superspace 
integrals,
\be
S\propto {\rm Im}\;\Big(\int {\rm d}^4x \;{\rm d}^4\theta\; 
F(W^I)\Big)\,,  \label{chiral}
\ee
where $F$ is is an arbitrary function of reduced chiral multiplets 
$W^I(x,\theta)$. Such multiplets carry the gauge-covariant degrees of 
freedom of a vector multiplet, consisting of a complex scalar 
$X^I$, a spinor doublet $\Omega^{iI}$, an anti-selfdual field-strength 
$F_{\m\n}^{-I}$ and a triplet of auxiliary fields $Y^I_{ij}$. 
This Lagrangian may coincide with the effective Lagrangian associated  
with some supersymmetric Yang-Mills theory, but for 
our purposes its origin is not relevant. To enable 
coupling to supergravity the holomorphic function must be 
homogeneous of second degree.  

The Lagrangian contains spin-1 kinetic terms proportional to 
\be
{\cal L}\propto i\Big( {\cal N}_{IJ}\,F_{\mu\nu}^{+I}F^{+\mu\nu J}\
-\ \bar{\cal N}_{IJ}\,F_{\mu\nu}^{-I} F^{-\mu\nu J} 
\Big)\,, \nonumber 
\ee
where $F_{\m\n}^{\pm I}$ are the (anti-)selfdual field strengths 
and $\cal N$ is proportional to the second derivative of the 
function $\bar F(\bar X)$. In addition there are moment couplings 
(to the fermions, or to certain background fields, to be 
discussed later), so that the field strengths $F^{\pm I}_{\m\n}$ 
couple linearly to tensors ${\cal O}^{\pm\m\n}_I$, whose form is 
left unspecified at the moment. Define 
\be
G^+_{\mu\nu I}={\cal N}_{IJ}F^{+J}_{\mu\nu} + {\cal O}_{\mu\nu 
I}^+\,, \label{defG}
\ee
and the corresponding anti-selfdual tensor that follows from 
complex conjugation, so that the field equations read 
$\partial^\mu \big(G_{\mu\nu I}^+ -G^-_{\m\n I}\big) =0$. The 
Bianchi identities and equations of motion   
for the Abelian gauge fields are invariant under the transformation
\be
\pmatrix{F^{\pm I}_{\mu\nu}\cr  G^\pm_{\mu\nu I}\cr} 
\longrightarrow  \pmatrix{U&Z\cr W&V\cr} \pmatrix{F^{\pm 
I}_{\mu\nu}\cr  G^\pm_{\mu\nu I}\cr}\,,\label{FGdual}
\ee
where $U^I_{\,J}$, $V_I^{\,J}$, $W_{IJ}$ and $Z^{IJ}$ are 
constant real  $(n+1)\times(n+1)$ submatrices.
{}From (\ref{defG}) and (\ref{FGdual}) one derives 
that $\cal N$ must transform as
\begin{equation}
\tilde{\cal N}_{IJ} = (V_I{}^K {\cal N}_{KL}+ W_{IL} )\,
\big[(U+ Z{\cal N})^{-1}\big]^L{}_J  \,.\label{nchange}
\end{equation}
To ensure that $\cal N$ remains a symmetric tensor, at 
least in the generic case, the transformation  
(\ref{FGdual}) must be an element of $Sp(2n+2,{\bf R})$  
(we disregard a uniform scale transformation). Owing to this 
restriction, the   
signature of the imaginary part of $\cal N$ is invariant. 
Furthermore the tensor $\cal O$ must change according to
\be
\tilde{\cal O}_{\mu\nu I}^+ = {\cal O}_{\mu\nu J}^+  \,[(U+Z{\cal 
N})^{-1}]^J{}_I\,. \label{ochange}
\ee
The required change of $\cal N$ is induced by a change of the scalar 
fields, implied by
\be
\pmatrix{X^{I}\cr  F_{I}\cr} \longrightarrow  \pmatrix{\tilde 
X^I\cr\tilde F_I\cr}=
\pmatrix{U&Z\cr W&V\cr} \pmatrix{X^{I}\cr  F_I\cr}\,. 
\label{transX}
\ee
The two transformations \eqn{FGdual} and \eqn{transX} are such 
that they precisely generate uniform symplectic rotations of the 
appropriate phase-space variables for the vectors and scalars 
of the $N=2$ vector multiplets, so that we are in fact dealing with  
canonical transformations. Also for the fermions, the symplectic 
transformations induce corresponding canonical transformations. 

In \eqn{transX} we included a change of $F_I$. Because the 
matrix is symplectic, one can show that 
the new quantities $\tilde F_I$ can be written as  
the derivatives of a new function $\tilde F(\tilde X)$. 
The new but equivalent set of equations of motion one obtains by 
means of the symplectic transformation (properly extended to other 
fields), follows from a Lagrangian based on $\tilde F$. It is 
possible to integrate \eqn{transX} and one finds  
\bea
\tilde F(\tilde X) &=& F(X)-{\textstyle{1\over2}}X^I F_I(X) 
\nonumber \\
&& + {\textstyle{1\over2}}\Big[ \big(U^{\rm T}W\big)_{IJ}X^I X^J 
+\big(U^{\rm T}V + W^{\rm T}  Z\big)_I{}^J X^IF_J
+ \big(Z^{\rm T}V\big){}^{IJ}F_I F_J \Big] \,, 
\label{newfunction}
\eea
up to a constant and terms linear in the $\tilde X^I$. In the coupling to 
supergravity, where the function must be homogeneous of second 
degree, such terms are obviously excluded\footnote{%
   The terms linear in $\tilde X$ in \eqn{newfunction} are 
   associated with constant translations in $\tilde 
   F_I$ in addition to the symplectic rotation shown in 
   \eqn{transX}. Likewise one may introduce constant shifts in
   $\tilde X^I$. Henceforth we ignore these shifts, which are 
   excluded for local supersymmetry, even in the presence of a 
   background. Constant contributions to $F(X)$ are always 
   irrelevant.}. %

The above expression \eqn{newfunction} is not always so useful, as it 
requires substituting $\tilde X^I$ in terms of $X^I$, or vice 
versa. When $F$ remains unchanged, $\tilde F(\tilde X) = F(\tilde 
X)$,  the theory is {\it invariant} under the 
corresponding transformations, but again this is hard to verify 
explicitly in this form. A more convenient method instead, is to verify 
that the substitution $X^ I\to \tilde X^I$ into the derivatives 
$F_I(X)$ correctly induces the symplectic transformations on the 
periods $(X^I,F_J)$. In the next section we present a few 
examples where the new function is determined explicitly.

Clearly $F(X)$ does not transform as a function under 
symplectic transformations and it is this aspect that is central 
to what follows. However, \eqn{newfunction} shows immediately 
that the combination  
\be
F(X)-{\textstyle{1\over2}}X^I F_I(X) \label{holfunct}
\ee
does transform as a function under the symplectic transformations, 
i.e., as $\tilde f(\tilde X) = f(X)$. There are more quantities 
that transform as functions under symplectic transformation, but 
they are usually not holomorphic. In the coupling to supergravity 
\eqn{holfunct} vanishes identically by virtue of the 
homogeneity of $F(X)$. It is no coincidence that in the context 
of the effective action of supersymmetric Yang-Mills theories, 
\eqn{holfunct} is often expressible in terms of the moduli (this 
happens whenever the $(X^I,F_J)$ satisfy certain Picard-Fuchs 
equations) and 
is therefore a function that must be invariant under the group 
of monodromy transformations, which is a subgroup of the 
symplectic group [6,12-14,11].

\section{An example}
\noindent{}To exhibit the effect of the symplectic 
reparametrizations consider the following example 
that decribes supergravity coupled to two vector multiplets. 
Hence $n=2$ and the symplectic transformations constitute the 
group $Sp(6;{\bf R})$. The holomorphic function is taken equal to 
\be
F(X)= -{X^1(X^2)^2\over X^0} \,,
\ee
and gives rise to the following K\"ahler potential, 
\be
K= -\ln (S+\bar S)(T+\bar T)^2\,, \quad \mbox{where} \quad 
iS=X^1/X^0,\quad iT=X^2/X^0 \,.  
\ee
In supergravity the quantities $X^I$ are sections of a complex 
line bundle, which can be parametrized in terms of the 
coordinates $S$ and $T$. The latter are the complex scalar fields 
belonging to the two vector multiplets. 

This example arises in the effective field theory corresponding 
to a compactification of the heterotic string on $K_3\times 
T^2$. Apart from the graviphoton there are three additional abelian 
vector fields whose scalar partners are the dilaton $S$ and the toroidal 
moduli $T$ and $U$. However, one of the vector multiplets has 
been frozen such that $U=T$. This example has been used to test 
certain consequences of string-string duality [15-17], 
following a proposal for dual pairs of $N=2$ string vacua in 
\cite{KV}.   

The corresponding K\"ahler manifold has $SU_{\rm S}(1,1)\times 
SU_{\rm T}(1,1)$ isometries. The first $SU(1,1)$ group correponds 
to $S$-duality and is generated by the symplectic matrices
$$
\pmatrix{ d\,&c&0&0&0&0\cr \noalign{\vskip1mm}
          b\,&a&0&0&0&0\cr \noalign{\vskip1mm}
          0\,&0&d&0&0&-{1\over 2}c\cr \noalign{\vskip1mm}
          0\,&0&0&a&-b&0\cr \noalign{\vskip1mm}
          0\,&0&0&-c&d&0\cr \noalign{\vskip1mm}
          0\,&0&-2b&0&0&a\cr}\,,
$$
where always $ad-bc=1$. This leads to the following 
transformations of the $X^I$,
$$
X^0\to dX^0+cX^1,\quad X^1\to bX^0+aX^1,\quad X^2 \to {X^2\over X^0} 
(dX^0+cX^1)\,,
$$
from which one determines the typical $SU(1,1)$ transformations
\be
iS\to {aiS+b\over ciS +d}\,,\qquad T\to T\,.
\ee
The second $SU(1,1)$ group corresponds to $T$-duality and is 
generated by the symplectic matrices
$$
\pmatrix{d^2&0&2cd&0&-c^2&0\cr \noalign{\vskip1mm}
         0&d^2&0&c^2&0&-cd\cr  \noalign{\vskip1mm}
         bd&0&ad+bc&0&-ac&0\cr \noalign{\vskip1mm}
         0&b^2&0&a^2&0&-ab\cr  \noalign{\vskip1mm}
         -b^2&0&-2ab&0&a^2&0\cr \noalign{\vskip1mm}
         0&-2bd&0&-2ac&0&ad+bc\cr}\,,
$$
where $a$, $b$, $c$ and $d$ now parametrize the second $SU(1,1)$ 
group and are again subject to $ad-bc=1$. These $T$-duality 
transformations give rise to  
\bea
X^0 &\to& {1\over X^0} (dX^0+cX^2)^2\,, \nonumber \\
X^1 &\to& {X^1\over (X^0)^2} (dX^0+cX^2)^2 \,,\\
X^2 &\to& {dX^0+cX^2\over X^0} (bX^0+aX^2)\,, \nonumber
\eea
thus leading to 
\be
S\to S\,,\qquad iT\to {aiT +b\over ciT +d}\,.
\ee

The above transformations constitute symmetries of the theory and 
therefore do not cause a change of the function $F(X)$. By this, 
we do {\it not} wish to imply that $F(X)$ is invariant under the above 
substitutions, but rather that the new function following from 
the symplectic reparametrization according to \eqn{newfunction}, 
coincides with the previous one.  

Of course, one may also consider symplectic transformations that 
do not have this property and therefore define a 
reparametrization rather than a symmetry. One such example is 
the symplectic rotation defined by
\be
\tilde X^2= \alpha F_2\,, \qquad \tilde F_2 = -{1\over \alpha} 
X^2 + \beta F_2\,, 
\ee
with all other fields unchanged. From this rotation we easily 
determine the submatrices $U$, $V$, $W$ and $Z$ of the symplectic 
matrix so that we can construct the new function according to 
\eqn{newfunction}. It is equal to  
\be
\tilde F(\tilde X) = {1\over 4\alpha^2} {\tilde X^0(\tilde X^2)^2\over
\tilde X^1} +{\beta\over 2\alpha} (\tilde X^2)^2 \,.
\ee
Another symplectic transformation, defined by 
\bea
\tilde X^0 &=& X^0\,, \qquad\quad \tilde F_0 = F_0\,,\nonumber \\
\tilde X^1 &=& \alpha_1 F_2\,, \qquad \,\tilde F_1 =- {1\over 
\alpha_1}X^2 +{\beta\over \alpha_1} F_1  
+\gamma F_2\,,  \\ 
\tilde X^2 &=& \alpha_2 F_1 \,, \qquad \,\tilde F_2 = -{1\over 
\alpha_2}X^1+\delta F_1 +  {\beta\over\alpha_2} F_2\,. \nonumber
\eea
leads to a new function that is qualitatively even more different, 
\be
\tilde F(\tilde X)= \pm {1\over \alpha_1\alpha_2}\sqrt{ -\alpha_2
\tilde X^0 (\tilde X^1)^2 \tilde X^2} +{\gamma\over
2\alpha_1}(\tilde X^1)^2 + {\beta\over \alpha_1\alpha_2}\tilde
X^1\tilde X^2 + {\delta\over 2\alpha_2} (\tilde X^2)^2\,.
\ee

Of course, the symmetry transformations in terms of the new 
coordinates are different. For instance, in the case above with 
$\alpha_1=\alpha_2=-1$ and $\beta=\gamma=\delta=0$, the 
symplectic matrices corresponding to $S$- and $T$-duality take 
the form  
$$
\pmatrix{d   & 0   & 0  & 0   & 0   & c \cr  \noalign{\vskip1mm}
         0   & a   & 0  & 0   & 2b  & 0 \cr  \noalign{\vskip1mm}
         0   & 0   & d  & c   & 0   & 0 \cr  \noalign{\vskip1mm}
         0   & 0   & b  & a   & 0   & 0 \cr  \noalign{\vskip1mm}
         0  &\ft12c& 0  & 0   & d   & 0 \cr  \noalign{\vskip1mm}
         b   & 0   & 0  & 0   & 0   & a \cr} \,, \quad
\pmatrix{d^2 & 0   & c^2  & 0   & 2cd & 0 \cr    \noalign{\vskip1mm}
         0   &ad+bc& 0    & 2ac & 0   & 2bd \cr  \noalign{\vskip1mm}
         b^2 & 0   & a^2  & 0   & 2ab & 0  \cr   \noalign{\vskip1mm}
         0   & ab  & 0    & a^2 & 0   & b^2 \cr  \noalign{\vskip1mm}
         bd  & 0   & ac   & 0   &ad+bc& 0  \cr   \noalign{\vskip1mm}
         0   & cd  & 0    & c^2 & 0   & d^2 \cr} \,.
$$

For certain symplectic rotations the transformation $X^I$ to 
$\tilde X^I$ is singular and in this case there is no new function 
$F$, although the $(\tilde X^I,\tilde F_I)$ still exist. The 
latter is merely a technical problem as the full Lagrangian can 
still be written down consistently in terms of the $(\tilde X^I,
\tilde F_I)$ and their derivatives \cite{Ceresole}. However, in 
the presence of charges one no longer has the possibility of 
performing arbitrary symplectic transformations. In the context 
of our example a symplectic rotation that does not lead to a new 
function $F$, is, for instance, 
\be
\hat X^1= F_1\,,\qquad \hat F_1=- X^1\,,
\ee
with the other fields unchanged. On the basis of the $(\tilde 
X^I,\tilde F_I)$, the $S$- and $T$-duality transformations 
are described by the following two symplectic matrices, respectively,
$$
\pmatrix{ ~&~&~&        0&-c&0\cr     \noalign{\vskip1mm}
          ~&d{\bf 1}&~& -c&0&0\cr     \noalign{\vskip1mm}
          ~&~&~& 0&0&-{1\over 2}c\cr  \noalign{\vskip1mm}
          0&-b&0&~&~&~\cr             \noalign{\vskip1mm}
          -b&0&0&~&a{\bf 1}&~\cr      \noalign{\vskip1mm}
          0&0&-2b&~&~&~\cr},
\; 
\pmatrix{d^2&-c^2&2cd&0&0&0\cr      \noalign{\vskip1mm}
         -b^2&a^2&-2ab&0&0&0\cr     \noalign{\vskip1mm}
         bd&-ac&ad+bc&0&0&0\cr      \noalign{\vskip1mm}
         0&0&0&a^2&-b^2&-ab\cr      \noalign{\vskip1mm}
         0&0&0&-c^2&d^2&cd\cr       \noalign{\vskip1mm}
         0&0&0&-2ac&2bd&ad+bc\cr}.
$$
In this basis the $S$-duality transformations leave the action 
invariant provided  we impose the restriction $c=0$. There is no 
symplectic basis  
in which $S$-duality is an invariance of the action. This follows 
from the fact that $SU_{\rm S}(1,1)$ is embedded into $Sp(6,{\bf 
R})$ according to ${\bf 2}\oplus{\bf 2}\oplus{\bf 2}$. On the 
other hand, the $T$-duality transformations leave the 
action manifestly invariant in this basis. This is 
possible, because $SU_{\rm T}(1,1)$ is embedded according to 
${\bf 3}\oplus{\bf 3}$. With these observations we conclude our 
discussion of the example.

\section{Symplectic covariance and holomorphy}
\noindent{}The symplectic transformations can also be performed 
in the presence of chiral background fields as well as in a conformal 
supergravity background.  To couple supersymmetric vector multiplets 
to (scalar) chiral background fields is straightforward as one 
can simply incorporate additional chiral fields $\Phi$ into the 
function $F$ that appears in the integrand of \eqn{chiral}.
Also the coupling to conformal supergravity is
known \cite{DWLVP}. We draw attention to the fact that the $W^I$ 
are reduced, while the $\Phi$ can be either reduced or general 
chiral fields.  

There are a number of situations where chiral 
backgrounds are relevant. In supersymmetric theories many of the 
parameters (coupling constants, masses) can be regarded as  
background fields that are frozen to constant values (so that 
supersymmetry is left intact). Because these background fields 
correspond to certain representations of supersymmetry, the way 
in which they appear in the theory -- usually both perturbatively 
as well as nonperturbatively -- is restricted by supersymmetry. 
In this way we may derive restrictions on the way in which 
parameters can appear. An example is, for instance, the coupling 
constant and $\theta$-angle of a supersymmetric gauge theory, 
which can be regarded as a chiral field frozen to a complex 
constant $iS= \theta/2\pi+i4\pi/g^2$. 
Supersymmetry now requires that the function $F(X)$ depends on 
$S$, but {\it not} on its complex conjugate. This strategy of 
introducing so-called {\it spurion} fields is not new. In the 
context of supersymmetry it has been used in, for instance, 
\cite{shifman,amati,seiberg} to derive nonrenormalization 
theorems and even exact results. 
                      
Spurion fields can also be used for mass terms of 
hypermultiplets. When considering  
the effective action after integrating out the hypermultiplets, 
the dependence on these mass parameters can be incorporated in 
chiral background fields. In this case the background fields must 
be restricted to reduced chiral fields. In the previous 
example this restriction was optional. 
On the other hand, it may also be advantageous to not restrict 
the background fields to constant values, in order to study 
an explicit breaking of supersymmetry \cite{GG,AGD}.

We add that this strategy of using background (spurion) fields is 
very natural from the point  
of view of string theory, where the moduli fields, which 
characterize the parameters of the (supersymmetric) low-energy 
physics, reside in supermultiplets. In heterotic $N=2$ 
compactifications the 
background field $S$ introduced above coincides with the complex 
dilaton field, which comprises the dilaton and the axion, and 
belongs to a vector multiplet. The dilaton acts as the 
loop-counting parameter for string perturbation theory. Although 
the full supermultiplet that contains the dilaton is now 
physical, the derivation of nonrenormalization theorems can proceed
in the same way \cite{nilles,DWKLL}. We should stress here that 
when restricting the background to a reduced chiral multiplet, 
one can just treat it as an additional (albeit external) 
vector multiplet. Under these circumstances one may consider 
extensions of the symplectic  
transformations that involve also the background itself. Of 
course, when freezing the background to constant values, one must 
restrict the symplectic transformations accordingly. The above 
strategy is especially useful when dealing with anomalous 
symmetries. By extending anomalous transformations to the 
background fields, the  variation of these fields can compensate 
for the anomaly. The extended non-anomalous symmetry becomes
again anomalous once the background is frozen to a contant 
value. This strategy can be advantageous when dealing with 
massive hypermultiplets.

Another context where chiral backgrounds are relevant concerns 
the coupling to the Weyl multiplet, which involves interactions 
of vector multiplets to the square of the Riemann tensor. In this 
case the scalar chiral background  
is not reduced and is proportional to the square of the Weyl 
multiplet. Here the strategy is not, of course, to freeze the 
background to a constant, but one is interested in more general 
couplings with conformal supergravity. In the context of string 
theory the coefficient functions 
in terms of the Weyl background were studied and 
evaluated from certain type-II string 
amplitudes \cite{AGNT1}. An intriguing feature is that these 
functions  
can be identified with the topological partition functions of a 
two-dimensional twisted nonlinear sigma model defined on a 
Calabi-Yau target space. These partition functions are obtained by 
appropriately integrating over  
genus-$g$ Riemann surfaces. However, they do not depend 
holomorphically on the Calabi-Yau moduli, but there is a 
holomorphic anomaly due to certain nongeneric 
contributions coming the boundary of the moduli space underlying 
the Riemann surfaces \cite{BCOV}. In string theory these can be understood 
from the propagation of massless states \cite{kaplu}. Here we 
study similar coefficient functions, but with respect to general chiral 
backgrounds, and derive very similar results by insisting on a 
certain behaviour under symplectic transformations. 

It is possible to perform the standard analysis of the symplectic 
reparametrizations in the presence of chiral background fields 
starting from functions $F$ that 
depend both on the gauge superfield strengths $W^I$ and on the background 
field $\Phi$ in a way that is a priori unrestricted. Then one can 
proceed exactly as before and examine the equivalence classes in 
the presence of the background. The transformation rules, 
however, will also depend on the background fields. This does 
not affect the derivation, but there are a number of new 
features. We assume the presence of a single chiral scalar 
background field (the generalization to more background fields is 
straightforward) whose lowest-dimensional bosonic  
component is denoted by $A$ (for 
details, see \cite{buckow}). It turns 
out that the symplectic reparametrizations are fully consistent in 
the chiral background. The function $F(X,A)$ still changes 
according to \eqn{newfunction}, where $A$ remains unaffected 
although the transformations themselves depend on $A$. As before, 
$F$ does not  
transform as a function under symplectic transformations, but it 
is possible to identify certain quantities that do transform in 
this way. One of them is, for instance, the K\"ahler potential, 
but the K\"ahler potential is never holomorphic. 
This lack of holomorphy is not an exception. There are very few 
quantities that transform as functions  
under symplectic transformations and are holomorphic at the same 
time. Two such functions are
$$
F(X,A)-\ft12 X^I F_I(X,A)\,, \quad \mbox{and} \quad F_A(X,A)\,,
$$
where $F_A$ denotes the first derivative of $F$ with respect to 
the background field $A$. All other symplectic functions that we 
have been able to identify, are not holomorphic. In particular, 
higher derivatives of $F(X,A)$ with respect to the background $A$ 
do not transform as functions under holomorphic parametrizations. 
This conclusion is rather disturbing when considering 
symplectic transformations that constitute an invariance. In that 
situation we have $\tilde F(\tilde X,A) = F(\tilde X, A)$, but in 
spite of that, the coefficient 
functions (proportional to multiple derivatives of $F(X,A)$ with 
respect to the  
background) are not invariant {\it functions} under the corresponding 
transformations. This is only so for the first derivative $F_A$. 

It turns out, however, that one can systematically modify the 
multiple-$A$ derivatives of $F$, such that  
they will transform as functions under symplectic transformations. 
Naturally such modified functions are expected to arise when 
evaluating the coefficient functions directly from 
some underlying theory, such as string theory. Here we should stress 
that, in the context of the Wilsonian action, the original 
(holomorphic) coefficient functions do not directly correspond 
to physically relevant quantities. Therefore they do not 
have to be invariant under the symmetries associated with a 
subgroup of the symplectic transformations (such as 
the target-space dualities in string theory). In the remainder of 
this section, we will be completely  
general and construct a hierarchy of modified coefficient 
functions transforming as functions under the symplectic group. 
Subsequently we derive the holomorphic anomaly equation 
pertaining to these functions. 

The construction of the modified multiple derivatives, which 
define the coefficient functions when expanding order-by-order in 
the background, proceeds as follows. First, assume that $G(X,A)$  
transforms as a function under symplectic transformations. Then 
one readily proves that also ${\cal D}G(X,A)$ transforms as a 
symplectic function, where
\be 
{\cal D}\,\equiv\, {\pa\over \pa A} +  
iF_{AI}N^{IJ}{\pa\over \pa X^J} \,,  \label{derivative} 
\ee
and 
$$ 
N_{IJ}\equiv 2\,{\rm Im} \,F_{IJ}\,, \qquad N^{IJ}\equiv 
\big[N^{-1}\big]^{IJ}\,. 
$$
We note that ${\cal D}$ and $N^{IJ}\pa_J$ commute. Using 
\eqn{derivative} one can directly  
write down a hierarchy of functions which are modifications of 
multiple derivatives $F_{A\cdots A}$,
\be
F^{(n)} (X, A) \,\equiv \, {1\over n!}{\cal D}^{n-1} F_A(X,
A)\, ,\label{Fn}
\ee
where we included an obvious normalization factor. 
All the $F^{(n)}$ transform as functions under symplectic 
functions. However, except for $F^{(1)}$, they  
are not holomorphic. The lack of holomorphy is governed 
by the following equation ($n>1$),
\be
{\pa F^{(n)} \over \pa \bar X^I}= \ft12 \bar F_I{}^{JK} 
\sum_{r=1}^{n-1}\; {\pa F^{(r)} \over \pa  X^J}\,{\pa F^{(n-r)} 
\over \pa  X^K}\,, \label{anomalyeq2}
\ee
where $\bar F_I{}^{JK}= \bar F_{ILM}\,N^{LJ}N^{MK}$. 
This equation resembles the equation for the holomorphic 
anomaly found in \cite{BCOV} for the topological partition 
functions of twisted Calabi-Yau nonlinear sigma models. The 
latter equation exhibits two terms, however, and only one of them 
coincides with the right-hand side of \eqn{anomalyeq2}. This is 
the term that  
arises from Riemann surfaces that tend to be pinched into two 
disconnected surfaces. The missing term corresponds to pinchings 
of a closed loop, which lowers the genus by one unit. In the 
context of our derivation the latter term is of a different 
nature than the first one, as the genus is tied to an expansion 
order-by-order in the background field, but it can probably be 
incorporated by making further modifications to the coefficient 
functions. The fact that only one term occurs in the above anomaly 
equation, implies that no integrability relation can be derived for 
$F^{(1)}$, which remains holomorphic here. 

We should stress that \eqn{anomalyeq2} was obtained in a 
very general context and applies to both rigid and local $N=2$ 
supersymmetry. In the latter case we have to convert to 
holomorphic sections $X^I(z)$ and also the $F^{(n)}$ can be 
regarded as sections of a complex line bundle. This requires to 
set $A=0$ in the coefficient functions, so that we have functions 
that are homogeneous in the $X^I$. For the conversion of 
\eqn{anomalyeq2} to the case of local supersymmetry, one may 
conveniently make use of the formula  
\be
N^{IJ} = e^{K} \Big[g^{A\bar B}\, (\partial_A+\partial_AK) X^I(z)\, 
(\partial_{\bar B}+\partial_{\bar B}K) \bar X^J(\bar z) - X^I(z)\,\bar 
X^J(\bar z) \Big]   \,,
\ee
where $K(z,\bar z)$ and $g_{A\bar B}(z,\bar z)$ are the K\"ahler 
potential and metric, respectively. However, there is more to the 
coefficient functions than their dependence on the coordinates 
$z$. In order to derive all the 
corresponding couplings one needs the full dependence on the 
sections $X^I(z)$, which is not encoded in the topological 
partition functions.  

The holomorphic anomaly can thus be viewed as 
arising from a conflict between the requirements of holomorphy 
and of a proper (covariant) behaviour under symplectic 
transformations. The nonholomorphic modifications exhibited above 
can be regarded as (part of) the threshold corrections that arise 
due to the propagation of massless states \cite{kaplu}. 
Although the modifications  
presented above are not unique (we can always add to them some other 
symplectic function), it seems that they are in some sense 
universal, at least in the context of a background expansion. This 
would explain why, on the one hand, they can be obtained in such a 
general setting, while, on the other hand, they precisely 
reproduce one of the terms of the anomaly equation 
obtained in a much more specific context \cite{BCOV}. We should  
also point out that there seems   
a certain similarity here with the philosophy taken in 
\cite{witten}, where the holomorphic anomaly of \cite{BCOV} is 
regarded as an obstruction to (naive) background 
independence, as the latter is related to the freedom of choosing 
a symplectic basis.

\vspace{5mm}
\par
\noindent{\bf Acknowledgements}
\par
\noindent{}I am grateful for valuable discussions with L. Alvarez-Gaum\'e, 
R. Dijkgraaf, A. Klemm, J. Louis, D. L\"ust, K. Sfetsos, S. Theisen and 
E. Verlinde. 


\end{document}